\documentclass[nofootinbib,twocolumn,preprintnumbers,amsmath,amssymb,appendix]{revtex4}
\usepackage[hyperfootnotes=true, linktocpage=true, citecolor=blue, linkcolor=blue, urlcolor=Maroon, filecolor=Maroon]{hyperref}
\usepackage[noconfig]{refstyle}
\usepackage{graphicx}
\usepackage{dcolumn}
\usepackage{bm}

\usepackage{xcolor}
\usepackage{multirow}

\usepackage{enumitem}
\newlist{todolist}{itemize}{2}
\setlist[todolist]{label=$\square$}
\usepackage{pifont}

\usepackage{tikz-cd}

\newcommand{\beq}{\begin{equation}}
\newcommand{\eeq}{\end{equation}}
\newcommand{\bea}{\begin{eqnarray}}
\newcommand{\eea}{\end{eqnarray}}
\newcommand{\mbb}{\mathbb}

\newcommand{\nn}{\nonumber}

\begin{document}


\title{Geometric Bounds on the 1-Form Gauge Sector}



\author{Seung-Joo Lee}
\affiliation{Center for Theoretical Physics of the Universe, Institute for Basic Science, Daejeon 34126, South Korea}
\email[]{seungjoolee at ibs.re.kr}

\author{Paul-Konstantin Oehlmann}
\affiliation{Department of Physics, Northeastern University, 
360 Huntington Avenue, Boston, MA 02115, United States}
\email[]{p.oehlmann at northeastern.edu}


\begin{abstract}\vskip 3mm\noindent
We classify the allowed structures of the discrete 1-form gauge sector in six-dimensional supergravity theories realized as F-theory compactifications. This provides upper bounds on the 1-form gauge factors $\mbb Z_m$ and in particular demands each cyclic factor to obey $m\leq 6$. Our bounds correspond to the universal geometric constraints on the torsion subgroup of the Mordell-Weil group of elliptic Calabi-Yau three-folds. For any F-theory vacua with at least one tensor multiplet, we derive the constraints from the $\mbb P^1$ fibration structure of the base two-fold and identify their physical origin in terms of the worldsheet symmetry of the associated effective heterotic string. The bounds are also extended to the F-theory vacua with no tensor multiplets via a specific deformation of the theory followed by a small instanton transition, along which the 1-form gauge sector is not reduced. We envision that our geometric bounds can be promoted to a swampland constraint on any six-dimensional gravitational theories with minimal supersymmetry and also extend them to four-dimensional F-theory vacua. 

 \end{abstract}


\maketitle



\section{Introduction and Summary}

The notion of symmetries has been significantly generalized in recent years. In particular, ordinary symmetries acting on local operators have been generalized to $p$-form symmetries with $p \geq 1$, under which non-local operators extended along $p$ spatial dimensions are charged~\cite{Gaiotto:2014kfa}. As various interesting field-theoretic constructions involve gauging of symmetries, it is of utmost importance to understand the gauging of generalized symmetries to reveal the full symmetry structure of a theory. 

Perhaps the simplest such generalization is the 1-form gauge symmetries. A prototypical example arises from the discrete 1-form symmetry,
\beq
\Gamma \subseteq Z(\hat G) \,,
\eeq
of a simply-connected non-abelian\footnote{In this note we focus on the part of the center symmetry originating purely from the non-abelian 0-form gauge sector if $\hat G$ acquires $U(1)$ factors. In presence of the latter, the quotient action $\Gamma$ may receive additional contributions from mixing of the center $U(1)$s, which we hope to address in future work.}  0-form gauge group $\hat G$, where $Z(\hat G)$ denotes the center of $\hat G$. Gauging of this 1-form symmetry $\Gamma$ is responsible for the global structure of the 0-form gauge sector, resulting in the non-simply-connected quotient,
\beq\label{Gamma}
G = \hat G/\Gamma\,,
\eeq
for the actual gauge group. 

If the gauge theory in scrutiny is coupled to gravity, such gauging of the 1-form symmetry $\Gamma$, or equivalently, taking the quotient~\eqref{Gamma} by $\Gamma$ of the 0-form gauge sector, is not an option but a must unless the symmetry is broken. This follows from the absence of global symmetries~\cite{Banks:2010zn, Harlow:2018tng}, which is one of the conjectured properties that any consistent quantum theories of gravity are subject to. In the {\it swampland program} initiated in~\cite{Vafa:2005ui}, many general criteria of this kind have been proposed as universal features of quantum gravity, dubbed {\it swampland constraints}, leading to invaluable inspirations as well as explicit guiding principles for a variety of recent research activities in high energy physics.  

The aim of this note is to propose general bounds on the discrete 1-form gauge symmetry $\Gamma$ of the form~\eqref{Gamma}  in six-dimensional gravitational theories with minimal supersymmetry, going beyond the necessity for its gauging. Mostly focusing for concreteness on F-theory vacua, we will constrain the structure of the 1-form gauge sector in 6d $\mathcal N=(1,0)$ supergravity theories as 
\beq
\Gamma = \mbb Z_n \times \mbb Z_m \,, 
\eeq
where the allowed values of $(n,m)$ with $n \leq m$ are restricted as 
\begin{align}\renewcommand{\arraystretch}{1.2}\label{thebounds}
\begin{array}{|c||c|c|c|}\hline
n & 1  & 2 &3 \\ \hline
~m~ & ~1, \ldots, 6~ & ~2,3,4~ & ~3~   \\ \hline
\end{array} \,\,. 
\end{align} 
This in particular imposes a definite upper bound,
\beq\label{bound:6}
m \leq 6 \,,
\eeq
on every cyclic factor $\mbb Z_m$. 

As will be reviewed in section~\ref{sec:review}, similar constraints were studied in~\cite{Cvetic:2020kuw} for the 8d $\mathcal N=1$ supergravity theories with 0-form gauge sector of rank $18$, where a few more cases with the following values of $(n,m)$ are also allowed, 
\begin{align}\renewcommand{\arraystretch}{1.2}\label{eq:MW2}
\begin{array}{|c||c|c|c|}\hline
n & 1  & 2 &4 \\ \hline
~m~ & ~7,8~ & ~6~ & ~4~   \\ \hline
\end{array} \,\,, 
\end{align} 
in addition to those in~\eqref{thebounds}.
This restriction can be understood in F-theory as an immediate consequence of the classification results on the elliptic K3 surfaces, whose field-theoretic derivation can also be given from the anomaly consideration alone.\footnote{The one-to-one correspondence between the elliptic K3 surfaces and the fully-consistent supergravity theories can be further sharpened by considering the worldvolume theories of probe 3-branes~\cite{Hamada:2021bbz}.}
It is thus intriguing to note that the constraints persist, even in an enhanced form, in lower dimensional theories with less supersymmetry. Also from the geometric point of view, the set of elliptic Calabi-Yau three-folds available for 6d F-theory vacua is much wilder than that of the elliptic K3 surfaces for 8d vacua, and hence, the extension of the geometric constraints on surfaces to those on three-folds naively looks surprising. 

In this regard, the connection between the 6d and the 8d vacua will be clarified in section~\ref{sec:F}, where we will constrain the 1-form gauge sector of 6d $\mathcal N=(1,0)$ supergravity theories in the framework of geometric F-theory compactifications to six dimensions. We will find that the generic $\mbb P^1$ fibration structure of the base two-fold will play a pivotal role. The only exceptional base lacking such a fibration is $\mbb P^2$, which leads to F-theory vacua with $T=0$ tensor multiplets. However, we will be able to deform those models and to make a transition to the models with $T>0$ tensor multiplets, without reducing the 1-form gauge sector along the way, thereby validating the derivation of the generic constraints even to the theories without tensor multiplets.

In principle, it still remains a logical possibility that the geometric bounds derived as such may not serve as a general swampland constraint. However, the validity of our proposal is bolstered as follows. Firstly, in section~\ref{sec:het} we will provide a rationale behind the bound~\eqref{bound:6} from the dual heterotic perspective. Our main message is that the $\Gamma$-quotient~\eqref{Gamma} must act non-trivially on the {\it perturbative} 0-form gauge sector of the dual heterotic string, whether or not it affects the non-perturbative sector.\footnote{The non-perturbative sector will also be affected in general by the quotient; see \cite{Dierigl:2020myk}.} One can already infer from this perturbative nature of~$\Gamma$ that the definite upper bound of 6 arises for each 1-form factor, which follows from the simple group theoretical fact that $\mbb Z_6$ is the maximal cyclic subgroup of $E_8$.\footnote{To be precise, this intuition is valid under the assumption that the perturbative sector in six dimension is embeddable into the worldsheet symmetry of the 10d heterotic string, as will be emphasized later.}  Such a perturbative nature of the 1-form gauge sector is reminiscent of that of the $U(1)$ 0-form gauge sector~\cite{Lee:2019skh}; we will emphasize this resemblance in section~\ref{sec:het}. 
Furthermore, inspired by how the F-theoretic geometry and the heterotic physics constrain the 1-form gauge sector, we will be naturally led in section~\ref{sec:App} to envision a concrete route for extending the bounds to general effective theories via decompactification to eight dimensions.  

In section~\ref{sec:App} we will in fact provide a number of applications and extensions of our findings. Upon addressing the aforementioned possibility of enhancing our string-theoretic bounds to swampland ones, we will connect the putative 6d F-theory vacua violating the bounds to certain infinite distance limits in the moduli space, which will also be argued to result in decompactification. Another natural extension to be discussed concerns the discrete 0-form gauge sector. We will argue that a cyclic 0-form gauge factor with order beyond $6$ cannot arise in 6d F-theory models with $T\geq 1$ tensor multiplets. We will then also conjecture that no matter states in those models can have a $U(1)$ charge larger than $6$. Moreover, our derivation of the six-dimensional bounds will also be extended to the {\it geometric} part of the 1-form gauge sector in 4d F-theory models, with a systematic analysis of the flux-induced sector left for future work.

\section{Background and Review}\label{sec:review}
A most versatile tool to date for classifying supergravity theories in $D=12-2d$ is F-theory compactifications on
elliptically fibred Calabi-Yau $d$-folds. In $D \geq 6$ dimensions, i.e., with $d \leq 3$, this allows us to define the effective theories in terms of algebraic
properties of the fibration alone, which can always be brought into a Weierstrass form 
\begin{align}
y^2=x^3 + f x z^4 + g z^6 \, .
\end{align}
Here, $z,x,y$ are the coordinates of the weighted projective space $\mathbb{P}^2_{[1:2:3]}$ and $(f, g)$ a pair of sections, 
\beq
f \in H^0(B_{d-1}, \bar K_{B_{d-1}}^{\otimes 4}) \,,\quad  g \in H^0(B_{d-1}, \bar K_{B_{d-1}}^{\otimes 6}) \,, 
\eeq
with $\bar K_{B_{d-1}}$ denoting the anti-canonical bundle of the F-theory base $B_{d-1}$. 

As is well-known, the $[p,q]$ 7-brane stacks are located in the codimension-one loci in the base where the discriminant $\Delta := 4f^3 + 27 g^2$ vanishes, and the worldvolume gauge algebra $\frak g_i$ carried by each brane stack is read off from the vanishing orders of $f$, $g$ and $\Delta$  via the classification of singular fibers by Kodaira and N\'eron. 
On the other hand, the refined information as to which gauge {\it group} $G_i$ arises from each stack is encoded in the torsional sections that generate the finite subgroup $\text{MW}(Y_d)_{\text{Tors}}$ of the Mordell-Weil group. In particular, it is known that the global structure of the total gauge group, 
\beq
G= \prod_i G_i \,,
\eeq
is determined as~\cite{Aspinwall:1998xj, Mayrhofer:2014opa}
\begin{align}
\pi_1 (G) = \text{MW}(Y_d)_{\text{Tors}}     \,,
\end{align}
which embeds into the total center of the simply-connected cover $\hat G$, 
\begin{align}\label{center-general}
Z(\hat G)= \prod_i \mathbb{Z}_{n_i} \, ,
\end{align}
via the embedding factors 
\begin{align}\label{embedding-general}
k_i  \in \mathbb{Z}_{n_i} \, .  
\end{align}

Geometrically, the allowed torsional structures, as well as the generic forms of the associated Weierstrass modes, have been systematically studied in~\cite{Aspinwall:1998xj}, leading to the realization of 
\beq
\text{MW}(Y_d)_{\text{Tors}}  = \mathbb{Z}_{n} \times \mathbb{Z}_m \,,
\eeq
with the possibilities given in \eqref{thebounds}. 
 Furthermore, it is known~\cite{Hajouji:2019vxs} that there are additional outliers with the values of $(n,m)$ given in~\eqref{eq:MW2}, which only apply
 to the cases with $d=2$, i.e., to K3 surfaces. 
 
One of the goals of this note is to derive these geometric constraints for Calabi-Yau three-folds in an explicit and constructive manner, and to understand them physically from the perspective of the dual heterotic string. In particular we will see, under a certain simplifying assumption, that 
the global gauge group structure must always embed into the perturbative $E_8 \times E_8$ (or $SO(32)$) current of the heterotic worldsheet, from which the constraints immediately follow. 

This is to be contrasted with the following independent, but less constructive route to the geometric bounds~\cite{Hajouji:2019vxs,Dierigl:2020lai}: In Type IIB string theory the axio-dilaton $\tau$ in general takes any values in the upper half-plane $\cal H$ modulo SL$(2,\mathbb{Z})$. Upon adding the point at infinity, the moduli space of $\tau$ is represented by the modular curve
\begin{align}
X(1) = \mathcal{H}^* / \text{SL}(2,\mathbb{Z})\, .
\end{align}
The full SL$(2,\mathbb{Z})$ group, however, does not preserve the $\mathbb{Z}_n\times \mathbb{Z}_m$ torsion
points but only the congruence subgroup\footnote{The relevant congruence subgroups are defined as 
\begin{align}
\Gamma_1(n)= \left\{ \gamma \in \text{SL}(2,\mathbb{Z}): \gamma \equiv \left(\begin{array}{cc}
1 & * \\ 0 & 1 
\end{array}\right) \text{ mod } n  \right\}     \, ,
\end{align} 
\begin{align}
\Gamma(n)= \left\{ \gamma \in \text{SL}(2,\mathbb{Z}): \gamma \equiv \left(\begin{array}{cc}
1 &  0 \\ 0 & 1 
\end{array}\right) \text{ mod } n  \right\}     \, .
\end{align}  
} $ \Gamma(n)\cap \Gamma_1(m) \subset \text{SL}(2,\mathbb{Z})$ does. The associated moduli space of $\tau$, i.e., its respective compactified modular curve, is given as
\begin{align} 
X(n,m)=  \mathcal{H}^* /  \Gamma(n) \cap \Gamma_1(m)  \, .
\end{align}
Those curves admit an intricate structure of cusps, of which $\tau$ values correspond to the singular fibers of Kodaira type $I_s$, in perfect match with those found in the generic Weierstrass models~\cite{Aspinwall:1998xj}.
  This structure is not unlike that of the gauging of an SL($2,\mathbb{Z})$ subgroup, i.e., taking an orbifold which leads to additional fixed points. Those fixed points are loci where D7 brane stacks reside as required by the discrete 1-form gauge symmetry \cite{Dierigl:2020lai}. For the torsion with large orders $(n,m)$, the modular curve
 $X(n,m)$ is sufficient to fix the full gauge group of the 8d F-theory vacuum.
 
While it was not claimed in~\cite{Aspinwall:1998xj} that~\eqref{thebounds} is the full list, the proof of its completeness has later been provided in~\cite{Hajouji:2019vxs}. In this note, we will find another, more constructive route to this classification result, manifesting the connection in geometry between elliptic K3 surfaces and elliptic Calabi-Yau three-folds.

\subsection{8d EFTs}\label{sec:review-8d}
8d $\mathcal N=1$ supergravity theories with ${\rm rk}(G) = 18$ have been characterized via F-theory in terms of elliptic K3 surfaces $Y_2$ over the base one-fold $B_1 = \mbb P^1$. In particular, the inequivalent fiber structures of $Y_2$, including the torsion subgroups of the Mordell-Weil groups, have been classified~\cite{Shimada}. 
Moreover, as pointed out in~\cite{Braun:2021sex}, MW$_{\text{Tors}}(Y_2)$ is equally well understood in terms of the non-primitive
embedding of the ADE singularities into the 
$\Lambda_{2,18}$ lattice of the elliptic K3 surfaces. 

All in all, the torsion subgroup, and hence, the 1-form gauge sector, takes the form,
\beq\label{Gamma-general}
\Gamma = \mbb Z_n \times \mbb Z_m\,,
\eeq
with the allowed values of $(n, m)$ given by \eqref{thebounds} and \eqref{eq:MW2}.
As emphasized in~\cite{Hajouji:2019vxs}, the outliers~\eqref{eq:MW2} do not apply to higher-dimensional fibrations $Y_{d>2}$, and the K3 surfaces realizing them are necessarily rigid, i.e., {\it extremal}, as the associated F-theory compactifications turn out to have the non-Abelian groups of ranks ${\rm rk}(G)=18$. 

In the mean time, as pointed out in~\cite{Cvetic:2020kuw} via a purely field-theoretic consideration, the 1-form symmetry might in general lead to a global anomaly. Absence of such an anomaly for the 1-form gauge sector thus imposes field-theoretic constraints on the embedding factors $k_i$~\eqref{embedding-general} in terms of the cyclic orders $n_i$ of the center~\eqref{center-general}, leading in turn to constraints on the 1-form gauge sector $\Gamma$. In particular, for each of the extremal outliers~\eqref{eq:MW2}, the 0-form gauge sector, including its global structure, turns out to be uniquely determined by such constraints: 
 e.g., for the 1-form gauge sector $\Gamma=\mbb Z_7$ with $(n,m)=(1,7)$ in the table~\eqref{eq:MW2}, the unique consistent 0-form gauge group is $G=SU(7)^3/\mathbb{Z}_7$, where the global structure is realized by the embedding factors,  
\begin{align}
\label{eq:Z7embedding}
(k_1, k_2, k_3) = (1,2,3) \,,
\end{align}
into the total center $Z(\hat G)=\mathbb{Z}_{7}^3$. 
 
Given such a concrete classification of consistent gauge groups, a natural question arises as to what would go wrong e.g. if $\Gamma$ acquired a cyclic factor $\mbb Z_m$ with $m \geq 9$. One way to address this is to invoke the orbifold picture of the fundamental domain of the axio-dilaton discussed before. The modular curve $X(1,m)$ for $m\geq 9$ introduces cusps corresponding to the singular fibers of Kodaira type $I_{s_i}$, which must thus be included in the geometry, leading to gauge groups with
\begin{align}
\text{rk}(G) =  \sum_i (s_i  -1 ) \geq 24  \, ,
\end{align}
where the inequality saturates for $m=9$. However, this exceeds the geometric upper bound of $18$, which follows from the Tate-Shioda-Wazir thoerem~\cite{shioda1, shioda2, wazir}, and hence, $m \geq 9$ cannot be realized in 8d F-theory vacua.

\subsection{6d EFTs}
F-theory has proven to be an invaluable geometric tool\footnote{In this note we focus on the geometric, i.e., unfrozen, phase of F-theory.} also for systematically exploring 6d $\mathcal N=(1,0)$ supergravity theories in terms of elliptic Calabi-Yau three-folds $Y_3$ over a base two-fold $B_2$.  
Compactness of the base $B_2$ is crucial as its finite volume relates to
the Planck mass
\begin{align}
M_{\text{Pl}}^{4}\propto \text{Vol}(B_2) \,, 
\end{align}
keeping the effective theory gravitational. 
Moreover, the gravitational and the gauge anomalies are beautifully reflected in the topological intersection structure of the base~\cite{Park:2011ji}. To begin with, the anomaly coefficient of each gauge algebra factor $\frak{g}_i$ is identified with the respective curve class
  $b_i$ of the discriminental component, with gauge coupling given as
 \begin{align}
 g^{-2}_i \propto \text{Vol}(b_i) \, .
 \end{align}
Furthermore, the anomaly cancellation via the Green-Schwarz-Sagnotti-West mechanism  \cite{Green:1984bx,Sagnotti:1992qw} follows from the geometry of the elliptic fibration $Y_3$: for instance, the matter spectrum of the fields charged under $\frak g_i = \frak{su}(N)$ carried by the curve of class $b_i$ with self-intersection $n$ and genus $g$ is fixed purely in terms of these topological data as\footnote{Symmetric representations can be obtained through singular divisors~\cite{Klevers:2017aku}: an adjoint representation there is anomaly equivalent to an symmetric and anti-symmetric representation.}
\begin{align}
\label{eq:SUNCond}
\left[(8-N) n+ 16 (1-g)\right] \mathbf{F}+ (n+1-g)\mathbf{ \Lambda} +  g\textbf{Adj} \,, 
\end{align} 
whose contribution to the anomalies can be shown to cancel with the intersection pairings between the anomaly coefficients.

As it turns out, not every six-dimensional supergravity theory fulfilling the anomaly criteria has an F-theoretic realization. This can be easily inferred from the fact that the set of distinct elliptic Calabi-Yau three-folds is finite~\cite{Gross, Grassi}, while infinite classes of such apparently-consistent supergravity theories arise (see, e.g.,~\cite{Taylor:2011wt}). It has recently been found~\cite{Kim:2019vuc}, however, that some of those infinite classes can be ruled out by the consistency requirement on certain BPS strings, whose existence is addressed by one of the swampland constraints known as the completeness hypothesis (see~\cite{Harlow:2018tng} for a recent account).  

In much the same way, the analogue of the field-theoretic 8d global anomalies by itself does not lead to as strong constraints on the 1-form gauge sector as the geometric ones~\eqref{thebounds}. The latter was first addressed in~\cite{Hajouji:2019vxs} in terms of the torsion subgroup of the Mordell-Weil group of $Y_3$ via modular curves. The precise argument there was based on the structure of the modular curves $X(n,m)$: the latter indicated that the generic Weierstrass models with ${\rm MW}(Y_3)_{\rm Tors} = \mbb Z_n \times \mbb Z_m$ involve collisions of singularities which effectively result in non-minimal codimension-one fibers with vanishing orders, 
\beq
({\rm ord}(f), {\rm ord}(g)) \geq (4,6) \,,
\eeq
and hence, should be ruled out. In the ensuing section, a constructive derivation will be given of the  same  geometric constraints. Furthermore, we will also provide in section~\ref{sec:het} a heterotic worldsheet perspective on those geometric constraints, thereby interpreting them physically, still in a string-theoretic context. In fact, as will be envisioned in section~\ref{sec:App}, it is our belief that the purely effective-{field}-theoretic constraints will eventually enhance, possibly to the geometric ones, once again by applying appropriate swampland constraints if necessary.

\section{Geometric Bounds from F-theory}\label{sec:F}
Let us consider 6d $\mathcal N=(1,0)$ supergravity theories realized as F-theory compactification on an elliptic Calabi-Yau three-fold, 
\beq\label{pi}
\pi: Y_3 \to B_2\,.
\eeq 
Upon requiring the absence of non-minimal fibers for $Y_3$ to admit a Calabi-Yau resolution, the base two-folds $B_2$ allowed by F-theory are restricted~\cite{Gross, Grassi} to be either\footnote{While the Enrique surfaces provide yet another viable base topology, we do not discuss them as they only lead to a trivial fibration.}
\beq\label{base-T=0}
B_2 = \mathbb P^2
\eeq
or 
\beq\label{base-T>0}
B_2={\rm Bl}^k (\mathbb F_a),
\eeq
a $k$-fold blowup of Hirzebruch surface $\mathbb F_a$. The F-theory vacua of these two types, respectively, have $T=0$ and $T=1+k \geq 1$ tensor multiplets. In this section we will derive geometric bounds on the 1-form gauge sector, starting from the theories of the latter type with $T \geq 1$.

\subsection{F-theory Vacua with $T \geq 1$}\label{sec:T>=1}

We start by noting that every base two-fold $B_2$ of the form~\eqref{base-T>0} admits a rational fibration, 
\beq\label{rho}
\rho: B_2 \rightarrow \mbb P^1\,,
\eeq
whose generic fiber $f$ is a rational curve.\footnote{The base $B_2$ is a Hirzebruch surface if $k=0$ in~\eqref{base-T>0}, in which case even the {\it general} fiber $f$ is rational.} Restricting the fibration~\eqref{pi} to this rational fiber, we then obtain an elliptic surface $Y_2 := \pi^{-1}(B_1)$ for $B_1 = f \simeq \mbb P^1$, 
\beq\label{p}
p: Y_2 \to B_1 \,.
\eeq 
The adjunction formula can then be exploited as follows to show that $Y_2$ is a K3 surface: 
\bea\nn
c_1(\bar K_{Y_2}) &=& c_1(\bar K_{Y_3}|_{Y_2}) - c_1(N_{Y_2/Y_3}) \\
&=& - p^*c_1(N_{B_1/B_2}) \\ \nn
&=& 0 \,, 
\eea
where we have used in the second line that $Y_3$ has a trivial first Chern class and that the fibration~\eqref{p} is a restriction of~\eqref{pi}, and in the third line that $B_1$ is a fiber of~\eqref{rho}.

Furthermore, it follows by construction that the Mordell-Weil groups of $Y_3$ and $Y_2$ satisfy the obvious inclusion relation, 
\beq\label{MW-incl}
{\rm MW}(Y_3) \subseteq {\rm MW}(Y_2) \,.
\eeq
This can be seen by observing that a given section $\sigma \in {\rm MW}(Y_3)$ restricts to $s \in {\rm MW}(Y_2)$ under the embedding of the two fibrations~\eqref{pi} and~\eqref{p} as 
\beq
\begin{tikzcd}[column sep=.2em]
\sigma: &
B_2 &\dashrightarrow &
Y_3
\\
s: &
  B_1 \arrow[u,hook] &\longrightarrow  &
  \,Y_2 \arrow[u,hook] 
\end{tikzcd}
\eeq
where the restricted section $s$ is holomorphic even when $\sigma$ is not since $B_1=f$ is a generic fiber of~\eqref{rho}, hence evading potential non-generic points in $B_2$ where the section $\sigma$ wraps a fibral curve.  

From the fact that $B_1$ is a generic fiber of $B_2$, it immediately follows that such a section restriction is one-to-one and importantly, upon restriction a torsional section remains torsional with the same order. In other words, the inclusion~\eqref{MW-incl} of the Mordell-Weil groups leads to
\beq\label{Tors-incl}
{\rm MW}(Y_3)_{\rm Tors} \subseteq {\rm MW}(Y_2)_{\rm Tors} \,,
\eeq
so that the torsional subgroup ${\rm MW}(Y_3)_{\rm Tors}$ of ${\rm MW}(Y_3)$ must take the form,
\beq
{\rm MW}(Y_3)_{\rm Tors} = \mbb Z_{n'} \times \mbb Z_{m'} \subseteq \mbb Z_n \times \mbb Z_m \,,
\eeq
where $(n,m)$ take values from~\eqref{thebounds} and \eqref{eq:MW2} that are realizable for an elliptic K3 surface. 
We therefore conclude that the bounds on the 1-form gauge sector of 8d supergravity theories persist in 6d F-theory vacua with $T \geq 1$.

However, all 1-form gauge symmetries viable in eight dimensions are not necessarily realizable also in six dimensions. Indeed the 6d bounds turn out to be stronger due to the outliers~\eqref{eq:MW2} that are known not to be realized as the structure of the Moredell-Weil torsion of $Y_3$~\cite{Hajouji:2019vxs}. In particular, $\mbb Z_{7}$ and $\mbb Z_8$ are not a viable 1-form gauge sector, which we will also support by switching to the dual heterotic frame in section~\ref{sec:het}.\footnote{In view of the explicit description for the fibration $Y_3$, as explained in section~\ref{sec:review-8d}, its elliptic K3 fiber with the torsional orders~\eqref{eq:MW2} is extremal and hence rigid, rendering the elliptic fibration within the K3 almost trivial in that $\tau$ is constant. One might naively think that extremal K3 surfaces can only be fibered over $T^2$. It is known~\cite{Kohl:2021rxy}, however, that non-trivial smooth fibrations with ${\frak{su}}(3)$ holonomy are realizable in certain rigid three-folds, where the K3 fibers may become reducible over points in the $\mathbb{P}^1$ base.
}

One may also attempt to rule out those outliers by combining field-theoretic and geometric constraints, independently of the rationale based on modular curves. In the following we will illustrate how this works with the example $\mbb Z_7$; only the key ideas will be presented here, with the full details relegated to Appendix~\ref{Appendix}. 

As reviewed in section~\ref{sec:review-8d}, the anomaly consideration leaves 
\beq\label{Z7-instance}
G=SU(7)^3/\mathbb{Z}_7\,,
\eeq
as a viable gauge group configuration in 8d supergravity theories. This still survives the analogous constraints in the dimensionally-reduced 6d theories.\footnote{Unlike in eight dimensions, field-theoretic consideration in six dimensions does not by itself guarantee that $G$ of the form~\eqref{Z7-instance} is the only allowed 0-form gauge symmetry with $\mbb Z_7$-quotient. However, thanks to the geometric constraints~\cite{Shimada} for the generic K3 fiber $Y_2$, we know that $\mbb Z_7$-quotient leads precisely to~\eqref{Z7-instance} for the perturbative 0-form gauge sector even in six dimensions. It thus suffices to rule this unique choice out to show that $\mbb Z_7$ is not a viable 6d 1-form gauge factor.} However, additional constraints arise in six dimensions: In order for the 1-form symmetry $\Gamma=\mbb Z_7$ to be gauged, the matter spectrum should lack certain representations that transform under the $\Gamma$ action. 
To be concrete, the fundamental, the anti-symmetric and the adjoint representations under each $SU(N)$ factor naturally appear in the spectrum as~\eqref{eq:SUNCond} and, respectively, have the center charges, 
 \beq
 q_{Z_N}(\mathbf{R}_r) = \frac1N, \frac2N, ~0 \quad (\text{mod}~1)\,.
 \eeq
One thus requires that the following $\Gamma$-charge of the representation ${\bf R} = ({\bf R}_{r_i})_{i=1}^3$ under $G$, 
 \begin{align}\label{van}
 \sum_i   q_{Z_{n_i}}(\mathbf{R}_{r_i}) \cdot k_i\,, 
\end{align}
should vanish modulo 1, if matter states in ${\bf R}$ are to be present in the spectrum at all. Here, the embedding factors $k_i$ are fixed as~\eqref{eq:Z7embedding} either via the same anomaly considertaion as in eight dimensions~\cite{Cvetic:2020kuw} or via the geometric classification results on the K3 surface fiber $Y_2$~\cite{Shimada}. Note that the $r_i$'s can take the values $0$ through $3$, where the trivial representation ${\bf R}_0 = \bf 1$ with vanishing center charge is included to systematically express the $\Gamma$-charge of $\bf R$. The vanishing of~\eqref{van} turns out to restrict the topology of the curve classes $b_i$ severely; one can in particular show that the genus $g_i$ of each 7-brane locus is strictly positive, 
\beq
g_i \geq 1\,,
\eeq
which in the end leads to the violation of the Calabi-Yau condition, in that the residual discriminant $\Delta'$ turns out to have its class,  
\beq\label{residual}
[\Delta']=[\Delta] - 7(b_1 + b_2 + b_3) =  \bar K_{B_{2}}^{\otimes 12} - 7(b_1 + b_2 + b_3) \,,
\eeq
lie outside of the Mori cone (see Appendix~\ref{Appendix} for the proof). We therefore conclude that $\mbb Z_7$ is ruled out as a 1-form gauge sector.   

As it turns out, we can use the arguments along precisely the same line to also rule out $\mbb Z_8$, thereby establishing in particular the definite upper bound of 6 for every cyclic 1-form gauge factor as in~\eqref{bound:6}. 

Having successfully ruled out $\mbb Z_7$ and $\mbb Z_8$, we are only left with two double-factor outliers in~\eqref{eq:MW2} to be further analyzed. The geometric classification results on the K3 surfaces~\cite{Shimada} again fix the embedding factors uniquely for each of these two remaining cases. Therefore, once again, the admissible representations are strongly constrained, suggesting that they may be ruled out along the same line.

\subsection{F-theory Vacua with $T=0$}\label{sec:T=0}

What remains to be analyzed is the geometry of the F-theory background $Y_3$ with base $B_2= \mbb P^2$, for which no tensor multiplets arise. In case the effective supergravity theories contained a strongly coupled conformal matter sector~\cite{DelZotto:2014hpa}, one could blow up the geometry at the location of the conformal matter in the base as
\beq\label{bireq}
\begin{tikzcd}[column sep=.2em]
\pi: &
Y_3 &\longrightarrow &
B_2
\\
\hat\pi: &
  \hat Y_3\arrow[u, ""] &\longrightarrow  &
  \,\hat B_2 \arrow[u,""] 
\end{tikzcd}
\eeq
which brings in a tensor multiplet to the theory. The Mordell-Weil torsion ${\rm MW}(\hat Y_3)_{\rm Tors}$ of the blown-up geometry $\hat Y_3$ would then be subject to the same constraints as those derived earlier in section~\ref{sec:T>=1}. Crucially, the Mordell-Weil group is a birational invariant and hence the pair of birationally equivalent fibrations $Y_3$ and $\hat Y_3$ in~\eqref{bireq} satisfy
\beq
{\rm MW}(\hat Y_3) \simeq {\rm MW}(Y_3) \,. 
\eeq
In particular, we have 
\beq
{\rm MW}(\hat Y_3)_{\rm Tors} \simeq {\rm MW}(Y_3)_{\rm Tors} \,, 
\eeq
indicating that the 1-form gauge sector of the effective theories with no tensor multiplets should also be subject to the constraints that apply to the theories with tensor multiplets, as long as conformal matter is present. 

In the rest of this section we will thus focus on the theories without conformal matter. Let us suppose that the fibration~\eqref{pi} has a non-trivial Mordell-Weil torsion ${\rm MW}(Y_3)_{\rm Tors}$ with an $m$-torsional generator. The generic Weierstrass models for such a torsional geometry have the sections $f \in H^0(\mbb P^2, {\cal O}_{\mbb P^2}(12))$ and $g \in H^0(\mbb P^2, {\cal O}_{\mbb P^2}(18))$ naturally constrained to take a particular polynomial form in terms of a collection of sections, 
\bea\nn
a_{i,\alpha_i} &\in& H^0(B_2, \bar K_{B_2}^{\otimes i} ) \\ \label{ai}
& =& H^0(\mbb P^2, {\cal O}_{\mbb P^2}(3i)) \,, \quad i=1, 2, \cdots \,,6  \,,
\eea
as worked out in~\cite{Aspinwall:1998xj} for a list of torsion subgroups.\footnote{Similar constraints arise even when ${\rm MW}(Y_3)_{\rm Tors}$ has  two generators, as also exemplified in the same reference.} Here, the second index $\alpha_i$ for each $i$ runs from 1 to $P_i \in \mbb Z_{\geq 0}$ with $P_i$ representing the number of degree-$3i$ sections, which can be $0$, $1$ or beyond. 
As an illustration, the ansatz derived for ${\rm MW}(Y_3)_{\rm Tors} = \mbb Z_2$ takes the following form: 
\bea
f&=&a_{4,1} - \frac13 a_{2,1}^2 \,,\nn \\ 
g&=&\frac{1}{27} a_{2,1} (2 a_{2,1}^2 - 9a_{4,1}) \,, \label{fg-Z2} \\
\Delta &=& a_{4,1}^2 (4 a_{4,1} - a_{2,1}^2) \,,\nn
\eea
where the discriminant $\Delta$ follows from the expressions for $f$ and $g$. 
In our notation, the values of $P_i$ in this example are
\beq
    P_i = 
\begin{cases}
    1~ & \text{if } i=2~\text{or}~4\\
    0~              & \text{otherwise} \,,
\end{cases} 
\eeq
and the generic Weierstrass models~\eqref{fg-Z2} are described by a total of $2$ sections $a_{2,1}$ and $a_{4,1}$ of the line bundles $\bar K_{B_2}^{\otimes 2}$ and $\bar K_{B_2}^{\otimes 4}$, respectively.

Let us denote by $P \,(\geq 1)$ the total number of sections~\eqref{ai} that appear in the polynomial ansatz for $f$ and $g$, i.e., 
\beq
P := \sum_{i=1}^6 P_i \,.
\eeq
If $P\geq 2$, we can split the set of $a_{i, \alpha_i}$ into a pair of complementary proper subsets and tune the sections by demanding common linear factors, e.g., $z_{I=0,1}$, raised to appropriate powers as
\beq\label{tune}
a_{i, \alpha_i} = z_I^i b_{i, \alpha_i} \quad \text{for}\quad I=0,1\,, 
\eeq
depending on which of the two subsets they belong to. Here, $z_{I=0,1}$ are part of the homogeneous coordinates $[z_0:z_1:z_2]$ of $\mbb P^2$ and $b_{i, \alpha_i} \in H^0(\mbb P^2, {\cal O}_{\mbb P^2}(2i))$ are generic sections.\footnote{Note that the generic choice of $b_{i,\alpha_i}$ in~\eqref{tune} could in principle result in a partial Higgsing of the non-Abelian 0-form symmetries in case the sections $a_{i, \alpha_i}$ had not been generic to begin with before the tuning. However, this does not affect our discussions about the bounds on the 1-form gauge sector.} It then follows that the Weierstrass sections $f$ and $g$ have vanishing orders
\beq\label{46}
({\rm ord}(f), {\rm ord}(g))= (4,6) 
\eeq
at the point $[0:0:1]$, while along each of the two individual loci $z_{I=0,1}=0$ at least one of the two orders is reduced. For instance, in the above example~\eqref{fg-Z2}, we have $P=2$ so that the tuning~\eqref{tune} goes as
\bea
a_{2,1} = z_0^2 b_{2,1} \,, \\ 
a_{4,1}= z_1^4 b_{4,1} \,,
\eea
which in turn leads to
\bea
f&=& z_1^4 b_{4,1} - \frac13 z_0^4 {b}_{2,1}^2 \,, \\ 
g&=& \frac{1}{27} z_0^2 (2 z_0^4  {b}_{2,1}^2- 9z_1^4 {b}_{4,1})\,,
\eea
exhibiting the codimension-two non-minimality at $[0:0:1]$ as expected by~\eqref{46}.

If $P=1$ on the other hand, $f$ and $g$ are expressible as a power of a unique section, 
\beq
a_{i_0} \in H^0(\mbb P^2, {\cal O}_{\mbb P^2}(3i_0))\,,
\eeq
where $i_0$ divides ${\rm gcd}(4,6)=2$, implying that $i_0=1$ or $2$. In the former case we have
\beq
f \propto a_1^4 \quad \text{and}\quad g \propto a_1^6 \,,
\eeq
leading to codimension-1 non-minimal fibers along the cubic locus $a_1=0$, which we may thus discard. In the latter case we do not have such non-minimal singularities as
\beq
f \propto a_2^2 \quad \text{and} \quad g\propto a_2^3 \,. 
\eeq
We may then tune the section $a_2$ to take the form,
\beq
a_2 = z_0 z_1 b_2 \,, \quad b_2 \in H^0(\mbb P^2, {\cal O}_{\mbb P^2}(4))\,,
\eeq
so that the theory develops a conformal matter sector at the point $[0:0:1]$.  

Therefore, whether $P\geq 2$ or $P=1$, the geometry can be tuned within the given $m$-torsional ansatz to develop an isolated codmension-two non-minimal fiber, allowing us to blow it up in the same fashion as in~\eqref{bireq}. The resulting effective theory thus acquires a tensor multiplet while the 1-form gauge sector remains intact under the blow up operation at least. Even though the prerequisite tuning operation might possibly have enhanced the 1-form gauge sector, the upper bounds obtained for theories with $T\geq 1$ would still apply to those with $T=0$.

\section{Heterotic Views on the Bounds}\label{sec:het}

The F-theoretic bounds obtained in the previous section hinges on the special geometric role played by the rational fiber $f$, which is a 0-curve in $B_2$.\footnote{Recall that for $B_2=\mbb P^2$, we have even enforced a codimension-2 non-minimal fiber to be able to blow $B_2$ up so that a 0-curve $f$ may be present in the blown-up base.} From the perspective of the effective theory as well, such a rational fiber $f$ in $B_2$ plays a distinguished physical role since D3 brane wrapped on it leads to an effective heterotic string. In fact such a realization of the heterotic string in F-theory has already played an important role in studying various swampland constraints such as the weak gravity conjectures~\cite{Lee:2018urn, Lee:2018spm, Lee:2019tst, Klaewer:2020lfg} and the swampland distance conjecture~\cite{Klaewer:2020lfg, Lee:2021qkx, Lee:2021usk}, and also in constraining the $U(1)$ 0-form gauge sector~\cite{Lee:2019skh}. In this section we will provide a clean physical intuition for the derived geometric bounds on the 1-form gauge sector in view of this heterotic string.   

The gauge sector of the 6d supergravity theory then naturally splits into the perturbative sector that is ``visible'' to the heterotic string, hence leading to worldsheet global symmetries, and the non-perturbative sector that is invisible on the string worldsheet. Geometrically, the former arises from the 7-brane loci that intersect with $f$ and the latter from the loci that do not, i.e., they are either a fibral or an exceptional $\mbb P^1$ in $B_2$. The inclusion relation~\eqref{Tors-incl} of the Mordell-Weil torsion then implies the effective theory cannot have (a proper subset of) the center 1-form symmetry of such a non-perturbative sector gauged by itself. In other words, every center gauging must involve the perturbative sector. 

The perturbative nature of the 1-form gauge symmetries is reminiscent of what is known about the $U(1)$ 0-form gauge symmetries. It was addressed in~\cite{Lee:2019skh} that the latter should embed into the current algebra of the perturbative heterotic string, based on which the number of $U(1)$ gauge factors was bounded. In geometric terms, this  led to the bold proposal  for a definite upper bound of $16$ for ${\rm rank}({\rm MW}(Y_3))$, which concerns the free part of the ${\rm MW}(Y_3)$. The current proposal about the 1-form gauge symmetries on the other hand concerns the torsion part of ${\rm MW}(Y_3)$ and demands in particular that each cyclic factor of ${\rm MW}(Y_3)_{\rm Tors}$ be a finite cyclic subgroup of $E_8 \times E_8$ or $SO(32)$.   Interestingly, the maximal such cyclic subgroup occurs for $E_8$ and is known to be $\mbb Z_6$ (see, e.g.,~\cite{Z6}), agreeing with the constraints derived in~\ref{sec:F}. 

At this point one may wonder how $\mbb Z_7$ and $\mbb Z_8$ could be realized as a 1-form gauge factor in 8d supergravity theories even if they are not a subgroup of $E_8$. This naively seems to contradict the developed heterotic perspective. 
However, a simple explanation comes from the fact that the internal space for the 8d heterotic string theory is simply a two-torus $T^2$.  The latter leads to a pair of Kaluza-Klein $U(1)$s that can in principle enhance to a non-abelian gauge group in a non-geometric background and also mix with the  10d-induced perturbative sector of maximal rank $16$.   
The perturbative 0-form gauge sector may thus enhance in eight dimensions to a rank-$18$ non-abelian group. 

Motivated by this eight-dimensional phenomenon, we must thus revisit the potential intricate enhancement of the 10d-induced perturbative sector to the 6d $\mathcal N=(1,0)$ supergravity theories. While imposing only the minimal supersymmetry sets the heterotic internal space as a K3 surface, there may still arise a pair of Kaluza-Klein $U(1)$s, e.g., for a singular geometry. Nevertheless, the indirect observation in~\cite{Hajouji:2019vxs} via modular curves that $\mbb Z_7$ and $\mbb Z_8$ cannot serve as a Mordell-Weil torsion informs us that such an enhanced 0-form sector cannot be quotiented by those cyclic groups. We are currently not aware of a simple reasoning based on heterotic worldsheet symmetry that can rule out $\mbb Z_{\geq 7}$ even when the mixing of the 10d-induced perturbative sector mixes with Kaluza-Klein $U(1)$s. In section~\ref{sec:App}, however, we will envision the possibility of having the 6d bulk supergravity theory decompactify to an 8d one at an appropriate limiting regime in the moduli space, where the field-theoretic consideration in~\cite{Cvetic:2020kuw} would eventually rule those high-order cyclic 1-form sector.   

 Going back to the simple situation where no such intricate mixing arises in the perturbative sector, the derived bound of 6 for each cyclic 1-form gauge factor is also consistent with the worldsheet current generated by the small instanton transitions. We exploited the latter in section~\ref{sec:T=0} to address the geometric bounds for F-theory vacua with $T=0$. The codimension-two non-minimal singularities with vanishing orders~\eqref{46} are well-known to correspond to the tensionless E-string, which exhibits an $E_8$ flavor symmetry. Depending on the way this string couples to the bulk gravitational theory, parts of the flavor symmetry are typically gauged. This gauging naturally constrains the representations, $U(1)$ charges and the dimension of the residual Higgs branch (see~\cite{Dierigl:2018nlv}). Thus, when performing such an E-string transition, we expect the string to couple to the global 1-form gauge group $\Gamma$ as well. However, such a (partial) gauging is only possible if $\Gamma$ is a subgroup of $E_8$, and hence $|\Gamma|$ cannot exceed $6$ in the same vein as the heterotic case. 

\section{Discussion and Applications}\label{sec:App}

In this note, the general upper bounds on the 1-form gauge sector of 8d supergravity theories have proven to apply also to F-theory compactifications to 6 dimensions. By focusing on the 6d F-theory vacua conserving precisely 8 real supercharges, we have in fact ruled out some of the high-order symmetries that are realizable in 8d supergravity, thereby proposing the following reduced set of viable 1-form gauge symmetries for 6d F-theory vacua: 
\bea  \label{Zk}
&\mathbb{Z}_{m} &\quad \text{for}\quad 2 \leq m \leq 6 \\ \label{Zkl}
&\mathbb{Z}_{n} \times \mathbb{Z}_{m} &\quad \text{for}\quad  (n,m) = (2,2), (2,4), (3,3)\,.
\eea

The geometric derivation of this result has hinged upon the distinguished role played by the ubiquitous rational fibration within the base two-fold $B_2$ other than $\mbb P^2$. As a key step we have identified the elliptic K3 surface $Y_2$ with the following property  
\beq\label{tors-incl}
{\rm MW}(Y_3)_{\rm Tors} \subseteq {\rm MW}(Y_2)_{\rm Tors}\,.
\eeq 
by restricting the elliptic fibration $Y_3$ over $B_2$ to the generic rational fiber of $B_2$. 
For F-theory vacua with $T \geq 1$ tensor multiplets, definite upper bounds on the 6d 1-form gauge sector have thus been established as the known geometric bounds on ${\rm MW}(Y_2)_{\rm Tors}$. Of the allowed set of Mordell-Weil torsions for K3 surfaces, the following four, 
\bea
&\mathbb{Z}_{m} &\quad \text{for}\quad m=7, 8 \,, \\  
&\mathbb{Z}_{n} \times \mathbb{Z}_{m} &\quad \text{for}\quad  (n,m) = (2,6), (4,4)\,,
\eea
have been argued not to be viable for ${\rm MW}(Y_3)_{\rm Tors}$ by also requiring absence of field-theoretic anomaly, while it is still possible to fiber the relevant rigid K3 surfaces non-trivially over $\mbb P^1$.

 With the 1-form gauge sector of the 6d F-theory vacua identified as ${\rm MW}(Y_3)_{\rm Tors}$, this leads to the candidates~\eqref{Zk} and~\eqref{Zkl} for 1-form gauge symmetries. On the other hand, for F-theory on elliptic $Y_3$ over $B_2=\mbb P^2$, it has been shown that the theory can be deformed to develop a conformal matter without reducing the 1-form gauge sector; since the tensor transition keeps the latter intact, the same upper bounds apply also to F-theory vacua without tensor multiplets.  

While the bounds~\eqref{Zk} and~\eqref{Zkl} originate from the geometric constraints on the internal space of F-theory, we have also provided a worldsheet interpretation of this result. That a central role is played by the rational fiber within $B_2$ strongly suggests that the physical object governing the geometric bounds must be the effective heterotic string; the latter can be identified, e.g., as in~\cite{Lee:2018urn, Lee:2018spm, Lee:2019tst, Klaewer:2020lfg, Lee:2019skh}, as a solitonic string arising from the $D3$ brane wrapped on the rational fiber. Indeed, we have been able to rederive the upper bound $6$ for the cyclic orders in~\eqref{Zk} and~\eqref{Zkl} from the fact that $\mbb Z_6$ is the maximal cyclic abelian subgroup of $E_8$, at least when the perturbative sector of the six-dimensional bulk theory does not acquire any intricate mixing with the Kaluza-Klein $U(1)$s. In such a simple situation, the validity of this heterotic interpretation has been addressed by showing that every discrete quotient action on the 0-form gauge sector necessarily invokes the perturbative sector. Interestingly, such a desired perturbative nature of the 1-form gauge sector has immediately followed from the inclusion relation~\eqref{tors-incl} derived in F-theory. We have also observed as an aside that the upper bound of $6$ is also consistent with the E-string transitions, where the relevant $\Gamma$-quotient needs to embed into the $E_8$ flavor group. 

To some extent the advocated perturbative nature of the 1-form gauge sector nicely parallels that of the $U(1)$ 0-form gauge sector of 6d F-theory vacua with $T \geq 1$. It was proven in~\cite{Lee:2019skh} that every $U(1)$ gauge factor, if present in the 0-form gauge sector of such theories, must embed into the perturbative current algebra of the effective heterotic string. Interestingly, the geometric origins of the $U(1)$ 0-form and the $1$-form gauge sectors are, respectively, the free and the torsional sectors of the Mordell-Weil group. The heterotic interpretation of the abelian 0-form gauge sector led in~\cite{Lee:2019skh} to the geometric conjecture that 
\beq\label{rk-bound}
{\rm rank}({\rm MW}(Y_3)) \leq 16\,,
\eeq
for an elliptic Calabi-Yau three-fold. In fact, already from the purely geometric perspective, the inclusion~\eqref{tors-incl} leads to a milder bound of $18$ since any elliptic K3 surface $Y_2$ obeys
\beq\label{rk-bound-Y2}
{\rm rank}({\rm MW}(Y_2)) \leq 18\,,
\eeq
which follows from the Tate-Shioda-Wazir thoerem, so that 
\beq\label{rk-bound-18}
{\rm rank}({\rm MW}(Y_3))  \leq {\rm rank}({\rm MW}(Y_2)) \leq 18 \,.
\eeq
There is a chance that the bound~\eqref{rk-bound} might possibly be violated since the rank-$16$ current algebra of the 10d heterotic string can in principle enhance in a compactified theory, for instance, by mixing with a pair of Kaluza-Klein $U(1)$s. 
Whether the milder upper bound~\eqref{rk-bound-18} could indeed be realized in an explicit 6d F-theory vacuum is yet to be clarified. 

 While both the 1-form and the $U(1)$ 0-form gauge sectors are perturbatively realized from the heterotic perspective, the addressed extension of the constraints~\eqref{Zk} and~\eqref{Zkl} on the former to F-theory models with $T=0$ does not have a natural counterpart on the latter. Those $T=0$ models can always be deformed to develop a conformal matter sector, which then lead to $T \geq 1$ models through a tensor transition. It was thus crucial that the deformation of our choice does not reduce the 1-form gauge sector, in arguing that the upper bounds on the 1-form sector of the latter models also bound that of the former. However, without generic ansatz available for $U(1)$ models in F-theory for a given positive rank, it is not clear how to find a systematic way to deform the models in a similar fashion, not reducing the $U(1)$ 0-form sector along the way. The rank bound~\eqref{rk-bound-18} thus remains conjectural for $T=0$ models.

Let us end with several extensions and applications of the current work, organized by the dimension $D$ of the spacetime in which the physics phenomena of interest take place.

\subsection{\uppercase{Decompactification $(D=8)$}}\label{sec:D=8}

\noindent{\bf Beyond the Geometric Bounds}

Our geometric bounds on the 1-form gauge sector in 6d F-theory have been built upon the analogous 8d bounds. However, there is a notable difference between the 6d and the 8d constraints: the latter can be derived by the absence of gauge anomalies, which provides a purely effective field theoretic perspective. One may then naturally attempt to also promote the geometric bounds in 6d to a general swampland constraint. The geometric arguments in this note suggest that one could achieve this by decompactifying the starting 6d ${\cal N}=(1,0)$ supergravity theory to an 8d supergravity endowed with the defect sector of a particular kind. Such a decompactification, if viewed as an F-theoretic deformation, corresponds to taking an infinite-distance limit in the K\"ahler moduli space, where the base $\mbb P^1$ of~\eqref{rho} expands while the fibral $\mbb P^1$ remains finite. One could thus analyze the obvious counterpart limit in supergravity to show that the resulting theory indeed decompactifies to begin with. Since this supergravity limit itself lies at infinite distance, it is natural to expect decompactification to occur, in line with the Emergent String Conjecture~\cite{Lee:2019wij}. Importantly, the 1-form gauge sector needs to be in good control along the way, as is the case for the prescribed 8d decompactification in F-theory. We leave it to future work to develop a general effective-theoretic picture for the desired, controlled decompactification. \\

\noindent{\bf Beyond the Minimal Fibers}

While the aforementioned decompactification would occur in F-theory at infinite distance in the K\"ahler moduli space, F-theory also provides a qualitatively distinct route to decompactification via complex structure limits, as initially analyzed for 8d F-theory vacua in~\cite{Lee:2021qkx, Lee:2021usk} and generalized to 6d vacua in~\cite{ALW}. 
In these inifnite-distance complex structure limits, several 7-brane stacks each carrying a 0-form Lie-type gauge algebra collide to enhance the symmetry to an emergent affine algebra. Geometrically, such an affine enhancement of the 0-form gauge algebra is realized by the non-minimal singularities in codimension-1 fibers of the elliptic Calabi-Yau manifold. In view of this, it is intriguing that the geometric bounds~\eqref{thebounds} and~\eqref{eq:MW2} for elliptic Calabi-Yau manifolds were understood in~\cite{Hajouji:2019vxs} as the orders of the cyclic groups beyond which codimension-2 non-minimal singularities at infinite distance arise. In other words, demanding that the 1-form gauge sector violate the bounds enforces the brane moduli to sit at infinite distance, resulting in a decompactification~\cite{ALW}. We leave a thorough investigation of this intriguing connection with infinite distance limits to future work. 

\subsection{\bf MIRROR $(D=6)$}

\noindent{\bf Bounds on Discrete 0-Form Gauge Factors}

An elliptically fibered Calabi-Yau three-fold $Y_3$ with a $n$-torsional section is conjectured\footnote{The mirror conjecture for fibrations is based on the mirror duality established for the elliptic curves and their generic structures. Strong evidence of the conjecture arises from explicit toric constructions~\cite{Klevers:2014bqa,Oehlmann:2016wsb} and more generally in the context of gauged linear sigma models~\cite{Schimannek:2021pau}. Also see~\cite{Berglund:1998ej,Cvetic:2015uwu} for the Heterotic/F-theory duality perspective.} to be paired with a genus-one fibered mirror dual three-fold $Z_3$ with a $n$-section. In view of the torsional bounds~\eqref{thebounds}, it is therefore conjectured that the $n$-section geometry $Z_3$ should also obey
\beq\label{ms-bound}
n \leq 6 \,.
\eeq
Explicit geometric constructions known to date of the multi-section geometry are consistent with the bound~\eqref{ms-bound}. The highest $n$ that has so far been achieved is $5$, which is realized via non-abelian gauged linear sigma models~\cite{Knapp:2021vkm} and attempts are made to construct explicit examples with a $6$-section. 

While the bound~\eqref{ms-bound} for multi-section geometry arises from the mirror conjecture, which is purely geometric and construction dependent in nature, the heterotic interpretation in section~\ref{sec:het} leads to an interesting string-theoretic intuition as follows. We first recall that compactifications of F-theory on $Y_3$ and $Z_3$ result in effective theories with $\mbb Z_n$ 1-form and 0-form gauge symmetries, respectively. In particular, the constraint~\eqref{ms-bound} is understood as the upper bound for the discrete $0$-form gauge symmetry $\mbb Z_n$. From the heterotic perspective, one can naturally obtain the latter by embedding\footnote{To be most general, we should consider embeddings of a gauge bundle into the full 10d heterotic current algebra potentially mixed with a pair of Kaluza-Klein U(1)s and rearranged afterwards. Here, we assume embeddings into $E_8$ only for simplicity of presentation; the physical intuition behind the mirror conjecture works in full generality.} into $E_8$ a gauge bundle with structure group $G$ of the form, 
\beq\label{G}
G = \hat G/\mbb Z_n \,\, \subset \,\, E_8  \,,
\eeq
where $\hat G$ is a simply-connected Lie group (or a product of such factors). This results in the 0-form gauge symmetry with a discrete $\mbb Z_n$ factor,  
\beq\label{H}
H= \mbb Z_n \times \hat H \,\, \subset \,\, E_8 \,
\eeq
where $\hat H$ is the commutant of $\hat G$ in $E_8$. To turn tables around, one can consider a gauge bundle with structure group $\hat H$ endowed with an $\mbb Z_n$-holonomy instanton to obtain $G$ as the effective gauge group, so that a discrete $\mbb Z_n$ 1-form symmetry is realized instead. 

The heterotic string thus provides us with the physical intuition behind the geometric mirror conjecture, pairing an elliptic Calabi-Yau three-fold $Y_3$ with a genus-one fibered $Z_3$. 
In particular we conclude from this dual heterotic perspective that the upper bound $6$ must apply not only to the 1-form but also to the 0-form factors, conforming with the geometric conjecture~\eqref{ms-bound}. 

Strictly speaking, however, the physical argument above is clear only for F-theory models with $T \geq 1$. It would be worth clarifying if this heterotic perspective extends to the models with $T=0$.  \\

\noindent{\bf Bounds on $U(1)$ Charges}

We emphasize that the aforementioned heterotic intuition behind the mirror conjecture relies heavily on the perturbative nature of the $1$-form gauge sector as addressed in section~\ref{sec:het}, as well as that of the discrete $0$-form gauge sector. The latter follows from the observation in~\cite{Lee:2019skh} that $U(1)$ 0-form gauge factors of 6d F-theory vacua should lie entirely in the perturbative sector. Presuming that a $\mbb Z_n$ 0-form gauge factor arises from the Higgsing of $U(1)$, the former should also be perturbative in nature.

In view of this Higgsing phenomenon, we are naturally lead to the closely related conjecture that the $U(1)$ charges $q$ of the matter states must be subject to 
\beq\label{q-bound}
q \leq 6 \,,
\eeq
for any F-theory models with $T \geq 1$. This conjecture is particularly interesting in view of recent attempts of classifying possible $U(1)$ charges in 6d F-theory (see e.g.~\cite{Baume:2015wia, Lawrie:2015hia, Cvetic:2015ioa, Taylor:2018khc, Raghuram:2018hjn, Collinucci:2019fnh}). 

While the conjectured charge bound~\eqref{q-bound} does not necessarily apply to the models on $B_2 = \mbb P^2$, it is particularly interesting to note that models on Hirzebruch surfaces were proposed in~\cite{Raghuram:2018hjn} to have matter states with a $U(1)$ charge as large as $21$. 
The apparent conflict between the conjectured bound~\eqref{q-bound} and the proposed models with such a large $U(1)$ charge would in principle be resolved by carefully analyzing the specific Higgsing processes assumed. On the one hand particular chains of Higgsings are required in~\cite{Raghuram:2018hjn}, both for non-Abelian and for (intermediate) multiple-$U(1)$ gauge theories; if any of them fail to be realized in F-theory the proposed models will not exist. On the other hand our bound~\eqref{q-bound} assumes that any matter state with a $U(1)$ charge $q$ could acquire a vev to break $U(1)$ to a $\mbb Z_q$ 0-form gauge factor. It will be interesting to clarify which Higgsing assumptions are invalid and also find the rationale behind such an obstruction, from the perspectives of both the geometry and the effective theory.\footnote{For instance, for the $U(1)$ Higgsing to be realizable in the effective theory, it is necessary that more than one Higgs fields exist and acquire a vev, so that D-flatness is satisfied. From this effective theory consideration alone, our conjectured bound~\eqref{ms-bound} on discrete 0-form gauge factor does not rule out any potential models in which precisely one massless field is charged under $U(1)$ with a large charge $q>6$. It is not clear, however, if such models can exist at all to begin with (see, e.g.,~\cite{Morrison:2021wuv}).}

\subsection{\uppercase{\bf compactification $(D=4)$}}
As already emphasized, the key to our geometric derivation of the bounds~\eqref{thebounds} is the presence of a $\mbb P^1$ fibration within the F-theory base $B_2$, with the exception of $B_2=\mbb P^2$, which, nevertheless, can be blown up without reducing the $1$-form gauge sector. In order to extend our results to 4d F-theory models, we must argue that $B_3$ also admits a $\mbb P^1$ fibration. Interestingly, a birational characterization of the F-theory base three-folds has recently been accomplished in terms of the so-called Fano towers of fibrations~\cite{fanotower1, fanotower2}. This suggests that the following {two exceptions} to a $\mbb P^1$ fibration should be analyzed: 
\begin{enumerate}
\item $B_3 = \mbb P^3$ 
\item $B_3 =$ a $\mbb P^2$ fibration over $\mbb P^1$ 
\end{enumerate}

In the former case, the same tuning used for $B_2 = \mbb P^2$ in section~\ref{sec:T=0} enables us to develop codimension-2 non-minimal fibers along a line, which, once blown up, will turn $B_3 = \mbb P^3$ into a $\mbb P^2$ fibration over $\mbb P^1$, which falls into the latter type. Similarly, the base $B_3$ of this latter type can be tuned within the torsional ansatz in a similar fashion, this time by choosing two toric coordinates of the $\mbb P^2$ fiber as the analogue of the linear factors in~\eqref{tune}. The common zero locus of these two toric divisors is a section of the $\mbb P^2$ fibration. Blowing up the three-fold $B_3$ along this section results in $\hat B_3$, which is a fibration of the Hirzebruch surfaces $\mbb F_1$, which may as well be interpreted as a $\mbb P^1$ fibration over yet another Hirzebruch surface. 
This assures that the torsional constraints~\eqref{thebounds} persist in the elliptic Calabi-Yau four-folds $Y_4 \to B_3$, thereby establishing the same bounds also on the 1-form gauge sector of 4d ${\cal N}=1$ F-theory vacua.   
 
Note, however, that this only constrains the 1-form gauge sector realized by the geometry alone. Unlike in 6d F-theory models, the 0-form gauge group of a 7-brane stack may be broken by gauge fluxes, which may in principle serve as an independent source of a 1-form gauge symmetry even for the non-perturbative sector, to which our heterotic argument based on $E_8$ embedding does not apply. 

Relatedly, it has recently been proposed~\cite{Li:2022vfj} that flux breaking of a non-abelian group may possibly lead to the Cartan $U(1)$s, a particular linear combination of which may admit a charged matter with charge as large as $657$, far beyond the torsional bound of $6$. Unless the matter fields with such a large $U(1)$ charge $q \gg 6$ face a fundamental obstruction in acquiring a vev, the $U(1)$ 0-form symmetry of the theory can be Higgsed to $\mbb Z_q$. In view of the mirror conjecture, this may serve as an indication for the existence of a 4d F-theory model with a $\mbb Z_q$ 1-form gauge factor, where $q$ violates the torsional bound. 

It would be interesting to extend our constraints on the perturbative 1-form gauge symmetry to such a non-perturbative sector.  

 \subsection{\uppercase{\bf decomposition $(D=2)$}}
Another interesting direction worth pursuing further arises from the worldsheet theories of the BPS non-critical strings
 that couple to bulk tensor fields. As was briefly mentioned in section~\ref{sec:review}, many of the seemingly-consistent configurations for the 0-form gauge sector can be successfully ruled out~\cite{Kim:2019vuc} by requiring the unitarity of the 2d superconformal field theories (SCFTs), to which the worldsheet theories of such BPS strings flow in the infra-red. 
However, this approach has not been strong enough to rule out all configurations known so far, which are believed to eventually be ruled out. Also, the consistency of the exceptional configurations such as $G=SU(7)^3/\mathbb{Z}_7$, which arise from the choices of the 1-form gauge sector with $(n,m)$ from~\eqref{eq:MW2}, remain agnostic from the unitarity consideration of such string probes alone.  

On the other hand, one might expect that the 1-form gauge sector descends to the 2d SCFTs as a global symmetry, as is the case for the 0-form sector~\cite{Dierigl:2022zll}. It is therefore worth questioning if any other worldsheet-theoretic notions, e.g., that of decomposition~\cite{Hellerman:2006zs}, which addresses the equivalence between $2$-dimensional quantum field theories with $1$-form global symmetries and disjoint unions of other theories, might possibly lead us to novel consistency constraints on the 6d bulk supergravity theories. It would be very interesting to clarify if and how the string worldsheet theories reflect the non-simply-connected nature of the 1-form global symmetry in a non-trivial fashion.

\section*{ACKNOWLEDGEMENTS} 
SJL is grateful to Timo Weigand for many important discussions as well as for invaluable comments on the manuscript. PKO would like to thank Markus Dierigl and Thorsten Schimannek for discussions and thanks the Cornell group for hospitality during the completion of this work.  
The work of SJL is supported by IBS under the project code, IBS-R018-D1. PKO has also received funding from the NSF CAREER grant PHY-1848089.

\appendix
\section{Inconsistency of $\mbb Z_7$ 1-Form Gauge Factor}\label{Appendix}
In this appendix we provide the detailed explanation on the inconsistency in geometry arising from demanding the 1-form gauge sector to take the form, 
\beq\label{Gamma-7-app}
\Gamma=\mbb Z_7 \,,
\eeq
in an F-theory compactification to six dimensions. Let us first recall from section~\ref{sec:T>=1} that the choice~\eqref{Gamma-7-app} made for the 1-form gauge sector uniquely fixes the 0-form gauge group $G$ as
\beq
G= SU(7)^3/\mbb Z_7 \,,
\eeq
which follows from the geometric classification of elliptic K3 surfaces~\cite{Shimada}. Here, each simply-connected factor, $\hat G_i = SU(7)$ for $i=1,2,3$, arises, in the language of the Calabi-Yau three-fold $Y_3$, from the singular fibers of Kodaira type $I_7$ along the curve in $B_2$ of class $b_i$. We denote the genus and self-intersection number of each such curve by $g_i$ and $n_i$, respectively. 
As explained around~\eqref{van}, the embedding factors $k_i$ are given as
\beq
(k_1, k_2, k_3)=(1,2,3) \,,
\eeq
so that the four representations $({\bf R}_{r})_{r=0}^4$ under each $SU(7)_i$ contribute to the $\Gamma$-charge, $q_\Gamma(\bf R)$, of ${\bf R} = ({\bf R}_{r_i})_{i=1}^3$ as 
\bea\nn
q_{Z_{n_1}}({\bf R}_{r})\cdot k_1 &=& 0,\frac17, \frac27, 0\,,\\ \label{contribution}
q_{Z_{n_2}}({\bf R}_{r})\cdot k_2 &=& 0,\frac27, \frac47, 0\,,\\ \nn
q_{Z_{n_3}}({\bf R}_{r})\cdot k_3 &=& 0,\frac37, \frac67, 0\,, 
\eea
if they occur in ${\bf R}$. Here, ${\bf R}_r$ for $r=0,1,2,3$ are, respectively, $\bf 1$, $\bf F$, $\bf \Lambda$ and $\bf Adj$ with the multiplicities, $\mu_{r=1,2,3}$, of the charged ones given as~\eqref{eq:SUNCond}. 

Suppose now that a given representation ${\bf R}$ appears in the spectrum. Then, its $\Gamma$-charge $q_\Gamma(\bf R)$ in~\eqref{van} necessarily vanishes modulo 1. In view of~\eqref{contribution}, we thus observe for a fixed $i$$^{\rm th}$ simple factor that if $\bf R$ transforms as $\bf R_{1}$ under $G_{i}$ then it should transform as $\bf R_2$ under some other $G_{i'}$. In other words, if $r_{i}=1$, then $r_{i'}=2$ for some $i' \neq i$. Similarly, it also follows that if $r_{i}=2$, then $r_{i'}=1$ for some $i'\neq i$. This simple observation then leads to the following constraints on the total multiplicities $\mu_1$ and $\mu_2$ for the given $i$-th sector: 
\bea \label{mu1}
\mu_1&=&n_{i} + 16(1-g_{i}) \overset{!}{=} 21 m_1 \,,\\ \label{mu2}
\mu_2 &=& n_{i}+1-g_{i} \overset{!}{=}  7 m_2 \,,
\eea
for $m_1, m_2 \in \mbb Z_{\geq 0}$, where we have used~\eqref{eq:SUNCond} for the multiplicities. Upon subtracting~\eqref{mu2} from~\eqref{mu1}, however, we obtain
\beq
15(1-g_{i}) = 7(3 m_1 - m_2) \,, 
\eeq
indicating that 
\beq
g_{i} \equiv 1~({\rm mod}~ 7)\,,
\eeq
and hence, in particular that 
\beq\label{>1}
g_{i} \geq 1\,.
\eeq

Let us now proceed to show that~\eqref{>1}, which should work for any choice of $i_0$, implies violation of the Calabi-Yau condition, in the sense that the class~\eqref{residual} of the residual discriminant $\Delta'$ is calculated not to be effective. This way, we will have concluded that $\mbb Z_7$ cannot be realized as a 1-form gauge sector. In the rest of this appendix, we thus prove this inconsistency by considering in turn the following three types of the base two-fold: 
\beq
B_2=\mbb P^2,\quad \mbb F_a,\quad {\rm and}\quad{\rm Bl}^k \mbb F_a\;{\rm with}\;\;k>0\,.
\eeq

\noindent{\bf Type 1: $B_2 = \mbb P^2$} 

The genus constraint~\eqref{>1} implies that the curve class $b_i$ is not too small, i.e., 
\beq
b_i = l_i \; H \,,\quad \text{with}~~l_i \geq 3\,,
\eeq
where $H$ is the hyperplane class. Since the $i^{\rm th}$ discriminantal component supports a singular fibers of Kodaira type $I_7$, the class of $[\Delta']$ is obeys 
\bea
[\Delta'] &=& \bar K_{B_2}^{\otimes 12}  - 7(b_1 + b_2 + b_3) \\ 
&=& (36-7(l_1+l_2+l_3)) H \\
&\leq & -27 H  \,,
\eea
in contradiction with the required effectiveness of $[\Delta']$. \\

\noindent{\bf {Type 2: $B_2 = \mbb F_a$}}

Let us denote the (irreducible) curve class $b_i$ as
\beq
b_i = s_i S + f_i F \,,   \quad s_i, f_i \geq 0\,,
\eeq
where $S$ and $F$ are the class of the exceptional section with self-intersection $-a$ and that of the fiber, respectively. We need to have the residual discriminant effective, whose class is computed as 
\bea
[\Delta'] &=& \bar K_{B_2}^{\otimes 12}  - 7(b_1 + b_2 + b_3) \\ 
&=& (24 - 7s)S+ (24+12a - 7f)F \,, 
\eea
where 
\beq\label{sf}
s:= s_1+s_2 +s_3\,,\quad f:=f_1+f_2+f_3\,. 
\eeq
Since $S$ and $F$ generate the Mori cone of $B_2$, it follows in particular that 
\beq\label{S-eff}
24-7s > 0 \,,
\eeq
and hence, 
\beq\label{0..3}
 0\leq s_i \leq 3\,, \quad \forall i\,.
\eeq
In the meantime one can calculate the genus $g_i$ via the Hirzebruch-Riemann-Roch theorem as
\beq\label{HRR}
2 g_i -2 = b_i \cdot (b_i - \bar K_{B_2})  \,,
\eeq
which leads to $g_i=0$ for $s_i=0$ and $1$. The genus constraint~\eqref{>1}, together with~\eqref{0..3} thus leave only two possibilities left: $s_i = 2$ or $3$. In particular, we thus have 
\beq
s_i \geq 2 \,, \quad \forall i \,,
\eeq
which, however, contradicts~\eqref{S-eff}.\\

\noindent{\bf {Type 3: $B_2 = {\rm Bl}^k \mbb F_a$ ($k>0$)}}

We can immediately deduce the inconsistency of this most general case from that of the Hirzebruch case. To this end, let us consider the morphism,
\beq
\nu: B_2 \to \mbb F_a \,,
\eeq
which blows down the exceptional divisors $E_l$ on $B_2$. The anti-canonical classes of $B_2$ and $\mbb F_a$ are naturally related by
\beq
\bar K_{B_2} = \nu^*(\bar K_{\mbb F_a}) + \sum_l c_l E_l \,,
\eeq
for the coefficients $c_l = -1$. 

Let us now denote the curve class $b_i$ as
\beq
b_i = s_i S + f_i F + \sum_l e_{i,l} E_l \,,
\eeq
where by abuse of notation we do not distinguish the classes $S$ and $F$ of the section and the fiber on $\mbb F_a$ from their pullbacks under $\nu$. Then, $[\Delta']$ is computed as
\bea
[\Delta'] &=& \bar K_{B_2}^{\otimes 12}  - 7(b_1 + b_2 + b_3) \\ \nn
&=& (24 - 7s)S+ (24+12a - 7f)F  \\   &&~+ \sum_l (12 c_l - 7e_l) E_l\,, 
\eea
where $e_l := e_{1,l} + e_{2,l} + e_{3,l}$ and $s$ and $f$ are again defined as~\eqref{sf}. 
Effectiveness of $[\Delta']$ implies that of its push-forward, 
\beq
\nu_*([\Delta']) = (24-7s)S + (24+12a-7f) F \,,
\eeq
so that we are still subject to~\eqref{S-eff} and hence, also to~\eqref{0..3}. Here, we have used $s_i \geq 0$, which follows from the requirement that $\nu_*(b_i) = s_i S + f_i F$ is effective on $\mbb F_a$. 
Invoking the Hirzebruch-Riemann-Roch theorem~\eqref{HRR}, we once again learn that $g_i=0$ for $s_i=0$ and $1$, and hence that $s_i \geq 2$, contradicting~\eqref{S-eff}.

\end{document}